\documentclass[english]{article}
\usepackage[T1]{fontenc}
\usepackage[latin9]{inputenc}
\usepackage{textcomp}
\usepackage{amsmath}
\usepackage{amsthm}
\usepackage{amssymb}
\usepackage{wasysym}

\makeatletter
\numberwithin{figure}{section}
\newcommand{\lyxaddress}[1]{
	\par {\raggedright #1
	\vspace{1.4em}
	\noindent\par}
}

\makeatother

\usepackage{babel}
\begin{document}
\title{\textbf{Schrödinger and Klein-Gordon theories of black holes from
the quantization of the Oppenheimer and Snyder gravitational collapse}}
\author{\textbf{Christian Corda}}
\maketitle

\lyxaddress{\textbf{SUNY Polytechnic Institute, 13502 Utica, New York, USA, Istituto
Livi, 59100 Prato, Tuscany, Italy and International Institute for
Applicable Mathematics and Information Sciences, B. M. Birla Science
Centre, Adarshnagar, Hyderabad 500063 (India)}}

\lyxaddress{\textbf{E-mail: }\textbf{\emph{cordac.galilei@gmail.com}}}
\begin{abstract}
The Schrödinger equation of the Schwarzschild black hole (BH) has
been recently derived by the Author and Collaborators. The BH is composed
of a particle, the \textquotedblleft electron\textquotedblright ,
interacting with a central field, the \textquotedblleft nucleus\textquotedblright .
Via de Broglie's hypothesis, one interprets the \textquotedblleft electron\textquotedblright{}
in terms of BH horizon's modes. Quantum gravity effects modify the
BH semi-classical structure at the Schwarzschild scale rather than
at the Planck scale. The analogy between this BH Schrödinger equation
and the Schrödinger equation of the s states of the hydrogen atom
permits us to solve the same equation. The quantum gravitational quantities
analogous of the fine structure constant and of the Rydberg constant
are not constants, but dynamical quantities having well defined discrete
spectra. The spectrum of the \textquotedblleft gravitational fine
structure constant\textquotedblright{} is the set of non-zero natural
numbers. Therefore, BHs are well defined quantum gravitational systems
obeying Schrödinger's theory: the \textquotedblleft gravitational
hydrogen atoms\textquotedblright . By identifying the potential energy
in the BH Schrödinger equation as being the gravitational energy of
a spherically symmetric shell, a different nature of the quantum BH
seems to surface. BHs are self-interacting, highly excited, spherically
symmetric, massive quantum shells generated by matter condensing on
the apparent horizon, concretely realizing the membrane paradigm.
The quantum BH descripted as a \textquotedblleft gravitational hydrogen
atom\textquotedblright{} is a fictitious mathematical representation
of the real, quantum BH, a quantum massive shell having as radius
the oscillating gravitational radius. Nontrivial consequences emerge
from this result: i) BHs have neither horizons nor singularities;
ii) there is neither information loss in BH evaporation, nor BH complementarity,
nor firewall paradox. These results are consistent with previous ones
by Hawking, Vaz, Mitra and others. Finally, the special relativistic
corrections to the BH Schrödinger equation give the BH Klein-Gordon
equation and the corresponding eigenvalues.
\end{abstract}

\section{Introduction}

It is a general conviction, which arises from the famous, pioneering
works of Bekenstein \cite{key-1} and Hawking \cite{key-2}, that
the role and the importance of BHs are fundamental in a quantum gravity
framework. BHs are indeed considered as being theoretical laboratories
for testing different models of quantum gravity. Bekenstein was the
first physicist who observed that, in some respects, BHs play the
same role in gravitation that atoms played in the nascent quantum
mechanics \cite{key-3}. This analogy implies that BH energy could
have a discrete spectrum \cite{key-3}. Therefore, BHs combine in
some sense both the ``hydrogen atom'' and the ``quasi-thermal''
emission in quantum gravity \cite{key-4}. As a consequence, BH quantization
could be the key to a quantum theory of gravity and, for that reason,
BH quantization became, and currently remains, one of the most important
research fields in theoretical physics of the last 50 years. Various
Authors proposed and still propose various different approaches. Hence,
the current literature is very rich, see for example {[}5\textendash 18{]}
and references within. 

An important longstanding problem in quantum gravity is the following:
what is the BH horizon and what happens at it? One finds various reasons
to take the BH horizon as a concrete place having real physical meaning.
The Bekenstein-Hawking entropy of the BH results proportional to the
area of the horizon \cite{key-19}. This permits to assign one bit
of information to each unit of area (in Planck units) of the BH horizon,
and, in principle, to quantize the BH horizon obtaining an integer
number of bits \cite{key-19}. In 1976 Hawking \cite{key-20} found
that if a BH horizon forms and if one assumes that effective field
theory concretely works away from such a BH horizon, then BH radiance
is obtained in a mixed state from the point of view of an observer
external to the BH and static with respect to it, while a freely falling
observer detects nothing unusual when crossing the BH horizon. This
is the BH information loss paradox. Among the various attempts to
resolve the paradox we recall the proposal that the BH horizon could
be described as a quantum surface having approximately one degree
of freedom per Planckian unit of area \cite{key-21}, the \emph{principle
of BH complementarity }\cite{key-22} and the \emph{firewalls} \cite{key-23}.
In 2014 Hawking \cite{key-24} proposed that BH event horizons could
not be the final result of the gravitational collapse. He hypotized
that the BH event horizon should be replaced by an \textquotedbl apparent
horizon\textquotedbl{} where infalling matter is suspended and then
released. Hawking did not give a mechanism for how this can work.
Hawking's conclusion was supported by Vaz, who found the cited mechanism
via entire solutions of the Wheeler-DeWitt equation, obtaining a compact
shell, that is, a dark star generated by matter condensing on the
apparent horizon during quantum collapse \cite{key-25}. In this paper
the same results of Hawking and Vaz will be obtained via a different
approach, that is Schrödinger theory of BHs, by showing that quantum
gravity effects modify the BH semi-classical structure not at the
Planck scale but at the Schwarzschild scale. This is in full agreement
with the results of Vaz in \cite{key-25}. On the other hand, before
the works of Hawking and Vaz, Mitra \cite{key-26}, Schild, Leiter
and Robertson \cite{key-27} and other Authors, see details in the
recent book of Mitra \cite{key-28}, proposed various approaches in
which the final result of the stellar gravitational collapse should
be an object having neither horizons nor singularities. A breakdown
of classical theory at the Schwarzschild scale is present also in
the fuzzball paradigm, which is a BH quantum description proposed
by string theory \cite{key-29,key-30}. But the final result of the
analysis in this paper is different with respect to such a fuzzball
paradigm where the quantum BH should be a sphere of strings with a
definite volume. In the approach of this paper it is instead shown
that the final result of the gravitational collapse is a self-interacting,
highly excited, spherically symmetric, quantum shell, a massive membrane
generated by matter condensing on the apparent horizon, which concretely
realizes the membrane paradigm. Hence, a series of remarkable consequences
arise from this result: i) BHs have neither horizons nor singularities;
ii) there is neither information loss in BH evaporation, nor BH complementarity,
nor firewall paradox.

\section{The quantum black hole: from Bohr to Schrödinger}

In the framework of quantum BHs, the Author and Collaborators developed
a semiclassical approach to BH quantization, see for example {[}31\textendash 33{]}
and the complete review \cite{key-34}, which is somewhat similar
to the historical semi-classical approach to the structure of a hydrogen
atom introduced by Bohr in 1913 \cite{key-35,key-36}. Recently, the
Author and Collaborators improved the analysis {[}37\textendash 39{]}.
An approach to quantization due to the famous collaborator of Einstein,
N. Rosen \cite{key-40} has been applied to the historical Oppenheimer
and Snyder gravitational collapse \cite{key-41}. Such an approach
shows that the BHs are the gravitational analog of hydrogen atoms
because they result as being composed by a particle, the ``electron'',
which interacts through a quantum gravitational interaction with a
central field, the ``nucleus''. This could, in principle, drive
to a space-time quantization based on a quantum mechanical particle-like
approach. Let us ask, what is the physical interpretation of the BH
``electron''? If one follows the analogy with the hydrogen atom,
the de Broglie hypothesis \cite{key-42} can be evoked, which enables
the wave nature of the BH \textquotedblleft electron\textquotedblright .
In other words, this ``particle'' does not orbit the nucleus in
an analogous way as a planet orbits the Sun. Instead, it should be
a standing wave representing a large and often oddly shaped \textquotedblleft atmosphere\textquotedblright{}
(the BH \textquotedblleft electron\textquotedblright ), which results
distributed around a central field (the BH \textquotedblleft nucleus\textquotedblright ).
Such an \textquotedblleft atmosphere\textquotedblright{} is interpreted
as being the BH horizon modes \cite{key-35,key-36}. The remarkable
consequence is, in turn, that quantum gravity effects modify the BH
semi-classical structure not at the Planck scale but at the Schwarzschild
scale. The framework where the radius of the event horizon shows quantum
oscillations was semi-classically introduced about 50 years ago \cite{key-43}
in terms of the so-called BH quasi-normal modes (QNMs). Such horizon
oscillations represent the BH back reaction due to external perturbations.
The absorptions of external particles as well as the emissions of
Hawking quanta are BH perturbations. The Bohr-like approach to BH
quantum physics {[}31\textendash 34{]} has been developed by the Author
and Collaborators starting from such an observation, by improving
previous works of Hod \cite{key-4} and Maggiore \cite{key-44}. The
QNMs framework is a semiclassical one similar to the approach that
Bohr developed in 1913 \cite{key-35,key-36} on the structure of the
hydrogen atom. But one also recalls that horizon modes have an importance
in quantum gravity frameworks if one considers them in terms of the
periodic motion of their particle-like analogue \cite{key-45}, by
evoking the de Broglie hypothesis. An energy spectrum which scales
as $\sim\sqrt{n}$ has been obtained in \cite{key-45}, and this seems
in agreement with the results in {[}31\textendash 34{]} and in {[}37\textendash 39{]}.
In the Bohr-like approach {[}31\textendash 34{]}, by considering the
absorptions of external particles, including the original BH formation,
and the emissions of Hawking quanta, the BH horizon is not fixed at
a constant radius from the BH \textquotedblleft nucleus\textquotedblright{}
\cite{key-33}. Due to energy conservation, the horizon contracts
during the emission of a particle and expands during an absorption
\cite{key-33}. Those quantum contractions/expansions must not be
considered as being ``one shot processes'' \cite{key-33}. Instead,
the horizon oscillates \cite{key-33}.

The results in {[}37\textendash 39{]} seems consistent the ones of
Hajicek and Kiefer \cite{key-11,key-12}, and permit to write down,
explicitly, the gravitational potential and the Schrödinger equation
for the BH as (hereafter Planck units will be used, i.e. $G=c=k_{B}=\hbar=\frac{1}{4\pi\epsilon_{0}}=1$)
{[}37\textendash 39{]}
\begin{equation}
V(r)=-\frac{M_{E}^{2}}{r},\label{eq: energia potenziale BH effettiva}
\end{equation}
\begin{equation}
-\frac{1}{2M_{E}}\left(\frac{\partial^{2}X}{\partial r^{2}}+\frac{2}{r}\frac{\partial X}{\partial r}\right)+VX=EX,\label{eq: Schrodinger equation BH effettiva}
\end{equation}
being $E=-\frac{M_{E}}{2}$ the BH total\emph{ energy} and \emph{$M_{E}$}
the \emph{BH effective mass} which has been introduced in BH physics
by the Author and Collaborators {[}31\textendash 34{]}. The BH\emph{
}effective mass is the average of the BH initial and final masses
which are involved in a quantum transition. \emph{$M_{E}$} indeed
represents the BH mass\emph{ during} the BH expansion (contraction),
which is triggered by an absorption (emission) of a particle {[}31\textendash 34{]}.
The rigorous definition of the BH\emph{ }effective mass is {[}31\textendash 34{]}
\begin{equation}
M_{E}\equiv M\pm\frac{\omega}{2},\label{eq: effective quantities absorption}
\end{equation}
being $\omega$ the mass-energy of the absorbed (emitted) particle
(the sign plus concerns absorptions, the sign minus concerns emissions).
Hence, one sees that introducing the BH effective mass in the BH dynamical
framework is very intuitive. But, for the sake of mathematical rigor,
that introduction has been completely justified via the Hawking's
periodicity argument \cite{key-46}, see {[}31\textendash 34{]} for
details. Therefore, in a quantum gravity framework the hole can be
interpreted as a particle, the ``electron'', which interacts with
a central field, the ``nucleus'' through the quantum gravitational
interaction of Eq. (\ref{eq: energia potenziale BH effettiva}) {[}31\textendash 34{]}.
One also observes that the Schrödinger equation for the BH of Eq.
(\ref{eq: Schrodinger equation BH effettiva}) is formally identical
to the traditional Schrödinger equation of the $s$ states ($l=0$)
of the hydrogen atom which obeys to the Coulombian potential \cite{key-47}
\begin{equation}
V(r)=-\frac{e^{2}}{r}.\label{eq: energia potenziale atomo idrogeno}
\end{equation}
The absence of angular dependence in Eq. (\ref{eq: Schrodinger equation BH effettiva})
makes it simpler than the corresponding Schrödinger equation for the
hydrogen atom from a point of view. From another point of view it
is more complicated instead. In fact, differently from Eq. (\ref{eq: energia potenziale atomo idrogeno}),
where the charge of the electron is constant, Eq. (\ref{eq: Schrodinger equation BH effettiva})
shows a variation of the BH mass. This is the physical reason for
the introduction of the BH\emph{ }effective mass, that is indeed a
dynamical quantity, in {[}31\textendash 34{]}.\textbf{ }In the recent
work \cite{key-39}, the Author and collaborators re-obtained the
BH Schrödinger equation (\ref{eq: Schrodinger equation BH effettiva})
via Feynman's path integral approach. In this context, the attentive
reader \cite{key-48} observes that despite a star is made of a very
large number of interacting particles (excitations of nonlinear quantum
fields), the dust star is here treated as a whole system rather than
a large number of dust particles, hence reducing the number of degrees
of freedom to just the star radius prior to quantisation. One might
thus wonder if the resulting quantum theory reduces to a simple hydrogen-like
problem because of this huge simplification - essentially the reduction
of many dust particle dynamics to an effective one-body problem \cite{key-48}.
In this case both Birkoff's theorem and no-hair theorem \cite{key-49}
come to our aid from the classical theory. Birkhoff's theorem \cite{key-49}
states that a spherical, nonrotating, BH must be the Schwarzschild
BH, which is also the final result of the Oppenheimer and Snyder gravitational
collapse above discussed. While the no-hair theorem states that the
Schwarzschild BH is characterized by only its mass (radius) \cite{key-49}.
Therefore it is really difficult to think that the uniqueness of the
\textquotedbl black hole\textquotedbl{} object that is obtained in
the classical theory can be lost when we pass to the quantum theory.
On the other hand we will see that further support for this conclusion
will be given later, in Section 4, when we see that the same Schrödinger
equation is also obtained from a non-perfectly homogeneous gravitational
collapse.

\section{Rigorous solution of the black hole Schrödinger equation: energy
spectrum and ``gravitational fine structure constant''}

One recalls that $e^{2}=\alpha$ is the fine structure constant, which,
in standard units, combines the constants $\frac{e^{2}}{4\pi\epsilon_{0}}$
from electromagnetism, $\hbar$ from quantum mechanics, and the speed
of light $c$ from relativity, into the dimensionlesso irrational
number $\alpha\simeq\frac{1}{137,036}$, which is one of the most
important numbers in Nature. From the present approach one argues
that the gravitational analogous of the fine structure constant is
not a constant. It is a dynamical quantity instead. As in natural
units the fine structure constant is exactly the squared electron
charge $e^{2}$, in the current case the charge is replaced by the
effective mass (the gravitational charge). Thus, calling the square
of the BH effective mass the \textquotedbl gravitational fine structure
constant\textquotedbl{} is due to this analogy. In fact, if one labels
the ``gravitational fine structure constant'' as $\alpha_{G},$
by confronting Eqs. (\ref{eq: energia potenziale BH effettiva}) and
(\ref{eq: energia potenziale atomo idrogeno}) one gets 
\begin{equation}
\alpha_{G}=\left(\frac{M_{E}}{m_{p}}\right)^{2},\label{eq: finezza gravitazionale}
\end{equation}
in standard units, where $m_{p}$ is the Planck mass. It will be indeed
shown that $\alpha_{G}$ has a spectrum of values which coincides
with the set of natural numbers $\mathbb{N}.$ In analogous way, the
Rydberg constant, $R_{\infty}=\frac{1}{2}m_{e}\alpha^{2},$ is defined
in terms of the electron mass $m_{e}$ and on the fine structure constant.
In the current gravitational approach the effective mass and the ``charge''
are the same. Hence, the quantum gravitational analogous of the Rydberg
constant is 
\begin{equation}
\left(R_{\infty}\right)_{G}=\frac{M_{E}^{5}}{2m_{p}^{5}l_{p}}.\label{eq: gravitational Rydberg}
\end{equation}
in standard units, where $l_{p}$ is the Planck length. Again, it
will be shown that $\left(R_{\infty}\right)_{G}$ is not a constant
but a dynamical quantity having a particular spectrum of values. 

By introducing the variable $y$
\begin{equation}
y(r)\equiv rX\label{eq: nuova variabile}
\end{equation}
Eq. (\ref{eq: Schrodinger equation BH effettiva}) becomes 
\begin{equation}
-\left(\frac{1}{2M_{E}}\frac{d^{2}}{dr^{2}}+\frac{M_{E}^{2}}{r}+E\right)y=0.\label{eq: Schrodinger nuova variabile}
\end{equation}
Setting $y'\equiv\frac{d}{dr}$ and using $E=-\frac{M_{E}}{2}$, Eq.
(\ref{eq: Schrodinger nuova variabile}) can be rewritten as 
\begin{equation}
y''+M_{E}^{2}\left(\frac{2M_{E}}{r}-1\right)y=0.\label{eq: Schrodinger riscritta}
\end{equation}
As it is $E<0,$ the asymptotic form of the solution, which is regular
at the origin, is a linear combination of exponentials $\exp\left(M_{E}r\right),$
$\exp\left(-M_{E}r\right)$. If one wants this solution to be an acceptable
eigensolution, the coefficient in front of $\exp\left(M_{E}r\right)$
must vanish. This happens only for certains discrete values of $E.$
Such values will be the energies of the discrete spectrum of the BH
and the corresponding wave function represents one of the possible
BH bound states.

If one makes the change of variable 
\begin{equation}
x=2M_{E}r\label{eq: cambio variabile}
\end{equation}
Eq. (\ref{eq: Schrodinger riscritta}) results equivalent to 
\begin{equation}
\left[\frac{d^{2}}{dx^{2}}+\frac{M_{E}^{2}}{x}-\frac{1}{4}\right]y=0,\label{eq: Schroedinger in x}
\end{equation}
and $y$ is the solution which goes as $x$ at the origin. For $x$
very large, it increases exponentially, except for certain particular
values of the BH effective mass where it behaves as $\exp\left(-\frac{x}{2}\right)$.
One wants to determine such special values and their corresponding
eigenfunctions. One starts to perform the change of function 
\begin{equation}
y=x\exp\left(-\frac{x}{2}\right)z\left(x\right),\label{eq: change of function}
\end{equation}
which changes Eq. (\ref{eq: Schroedinger in x}) to 
\begin{equation}
\left[x\frac{d^{2}}{dx^{2}}+\left(2-x\right)\frac{d}{dx}-\left(1-M_{E}^{2}\right)\right]z=0.\label{eq: Schroedinger con z}
\end{equation}
This last equation is a Laplace-like equation. Within a constant,
one finds only a solution which is positive at the origin. All the
other solutions have a singularity in $x^{-1}.$ One can show that
this solution is the confluent hypergeometric series 
\begin{equation}
Z=\sum_{i=1}^{\infty}\frac{\varGamma\left(1+i-M_{E}^{2}\right)}{\varGamma\left(1-M_{E}^{2}\right)}\frac{1}{\left(1+i\right)!}\frac{x^{i}}{i!}.\label{eq: hypergeometric series}
\end{equation}
In fact, one expands the solution of Eq. (\ref{eq: Schroedinger con z})
in Mc Laurin series at the origin as 
\begin{equation}
z=1+\alpha_{1}x+\alpha_{2}x^{2}+...\alpha_{i}x^{i}+...\label{eq: Mc Laurin}
\end{equation}
By inserting Eq. (\ref{eq: Mc Laurin}) in Eq. (\ref{eq: Schroedinger con z})
one writes the LHS in terms of a power series of $x$. All the coefficients
of this expansion must be null. Thus, 
\begin{equation}
\begin{array}{c}
2\alpha_{1}=1-M_{E}^{2}\\
\\
2*3\alpha_{2}=\left(2-M_{E}^{2}\right)\alpha_{1}\\
\\
....\\
\\
i\left(1+i\right)\alpha_{i}=\left(i-M_{E}^{2}\right)\alpha_{i-1}.
\end{array}\label{eq: coefficienti}
\end{equation}
Then, one gets 
\begin{equation}
\alpha_{i}=\frac{\left(i-M_{E}^{2}\right)}{\left(1+i\right)}\frac{\left(i-1-M_{E}^{2}\right)}{\left(1+i-1\right)}...\frac{1-M_{E}^{2}}{1+1}\frac{1}{i!},\label{eq: alfa i}
\end{equation}
which is exactly the coefficient of $x^{i}$ in the expansion (\ref{eq: Mc Laurin}). 

From a mathematical point of view the series (\ref{eq: hypergeometric series})
is infinite and behaves as 
\[
\frac{\exp x}{x^{\left(1+M_{E}^{2}\right)}}
\]
for large $x.$ Consequently, $y$ behaves in the asymptotic region
as 
\[
\frac{\exp\frac{x}{2}}{x^{M_{E}^{2}}}
\]
and cannot be, in general, an eigensolution. On the other hand, for
particular values of $M_{E}^{2}$ all the coefficients will vanish
from a certain order on. In that case, the series (\ref{eq: hypergeometric series})
reduces to a polynomial. The requested condition is 
\begin{equation}
1-M_{E}^{2}\leq0,\:with\:\left|1-M_{E}^{2}\right|\in\mathbb{N},\label{eq: condition}
\end{equation}
which implies the quantization condition 
\begin{equation}
M_{E}^{2}=n=n'+1,\:with\:n'=0,1,2,3,....+\infty.\label{eq: quantizzazione}
\end{equation}
The hypergeometric series becomes a polynomial of degree $n'$ and
$y$ behaves in the asymptotic region as 
\[
\frac{x^{n}}{\exp\frac{x}{2}}.
\]
Then, the regular solution of the BH Schrödinger equation becomes
an acceptable eigensolution. The quantization condition (\ref{eq: quantizzazione})
permits us to obtain the spectrum of the effective mass as 
\begin{equation}
\left(M_{E}\right)_{n}=\sqrt{n},\:with\:n=1,2,3,....+\infty.\label{eq:spettro effettivo}
\end{equation}
The quantization condition (\ref{eq: quantizzazione}) gives also
the spectrum of the ``gravitational fine structure constant'' that
one writes as 
\begin{equation}
\left(\alpha_{G}\right)_{n}=\left[\frac{\left(M_{E}\right)_{n}}{m_{p}}\right]^{2}=n,\:with\:n=1,2,3,....+\infty.\label{eq: struttura fine gravitazionale}
\end{equation}
Thus, the gravitational analogous of the fine structure constant is
not a constant. It is a dynamical quantity instead, which has the
spectrum of values (\ref{eq: struttura fine gravitazionale}) which
coincides with the set of non-zero natural numbers $\mathbb{N}-\left\{ 0\right\} .$
In the same way, one finds the spectrum of the ``gravitational Rydberg
constant'' as 
\begin{equation}
\left[\left(R_{\infty}\right)_{G}\right]_{n}=\frac{\left(M_{E}\right)_{n}^{5}}{2m_{p}^{5}l_{p}}=\frac{n^{\frac{5}{2}}}{2},\:with\:n=1,2,3,....+\infty.\label{eq: Rydberg gravitazionale}
\end{equation}
Now, from the quantum point of view, one wants to obtain the mass
eigenvalues as being absorptions starting from the BH formation, that
is from the BH having null mass, where with ``the BH having null
mass'' one means the situation of the gravitational collapse before
the formation of the first event horizon. This implies that one must
replace $M\rightarrow0$ and $\omega\rightarrow M$ in Eq. (\ref{eq: effective quantities absorption}).
Thus, one gets
\begin{equation}
M_{E}\equiv\frac{M}{2}.\label{eq: effective quantities absorption finali}
\end{equation}
By combining Eqs. (\ref{eq:spettro effettivo}) and (\ref{eq: effective quantities absorption finali})
one immediately gets the BH mass spectrum as 
\begin{equation}
M_{n}=2\sqrt{n},\label{eq: spettro massa BH finale}
\end{equation}
which is the same result of {[}37\textendash 39{]}. Eq. (\ref{eq: spettro massa BH finale})
is consistent with the BH mass spectrum found by Bekenstein in 1974
\cite{key-57}. Bekenstein indeed obtained $M_{n}=\sqrt{\frac{n}{2}}$
by using the Bohr-Sommerfeld quantization condition because he argued
that the BH behaves as an adiabatic invariant. It is also consistent
with other BH mass spectra in the literature, see for example. The
relationship between the BH effective mass and the BH total energy
is {[}37\textendash 39{]}
\begin{equation}
E=-\frac{M_{E}}{2},\label{eq: energia totale reale}
\end{equation}
which permits to find the BH energy spectrum as
\begin{equation}
E_{n}=-\sqrt{\frac{n}{4}}.\label{eq: BH energy levels finale.}
\end{equation}
The (unnormalized) wave function corresponding to each energy level
is 
\begin{equation}
X(r)=\frac{1}{r}\left[\frac{2M_{E}r}{\exp\frac{x}{2}}\left(\sum_{i=0}^{n-1}\left(-\right)^{i}\frac{\left(n-1\right)!}{\left(n-1-i\right)!\left(i+1\right)!i!}\right)\left(2M_{E}r\right)^{i}\right].\label{eq: wave function}
\end{equation}
One observes that the number of nodes of the wave function is exactly
$n-1$ and that the energy spectrum (\ref{eq: BH energy levels finale.})
contains a denumerably infinite number of levels because $n$ can
take all the infinite values $n\in\mathbb{N}-\left\{ 0\right\} .$
One can calculate the energy jump between two neighboring level as
\begin{equation}
\triangle E=E_{n+1}-E_{n}=\sqrt{\frac{n}{4}}-\sqrt{\frac{n+1}{4}}=-\frac{1}{4\left(\sqrt{\frac{n}{4}}+\sqrt{\frac{n+1}{4}}\right)}.\label{eq: jump}
\end{equation}
For large $n$ one obtains 
\[
\triangle E\simeq-\frac{1}{4\sqrt{n}}
\]
when $n\rightarrow+\infty$ one gets $\triangle E\rightarrow0.$ Thus,
one finds that the energy levels become more and more closely spaced
and their difference tends to $\triangle E=0$ at the limit, at which
point the continuous spectrum of the BH begins.

One also observes that a two particle Hamiltonian
\begin{equation}
H\left(\vec{p},r\right)=\frac{p^{2}}{2M_{E}}-\frac{M_{E}^{2}}{r},\label{eq: Hamiltonian}
\end{equation}
which governs the BH quantum mechanics, must exist in correspondence
of Eqs. (\ref{eq: energia potenziale BH effettiva}) and (\ref{eq: Schrodinger equation BH effettiva}).
Therefore, the square of the wave function (\ref{eq: wave function})
must be interpreted as the probability density of a single particle
in a finite volume. Thus, the integral over the entire volume must
be normalized to unity as 
\begin{equation}
\int dx^{3}\left|X\right|^{2}=1.\label{eq: Normalizzazione}
\end{equation}
For stable particles, this normalization must remain the same at all
times of the BH evolution. This issue has deep implications for the
BH information paradox \cite{key-20} because it guarantees preservation
of quantum information. As the wave function (\ref{eq: wave function})
obeys the BH Schrödinger equation (\ref{eq: Schrodinger equation BH effettiva}),
this is assured if and only if the Hamiltonian operator (\ref{eq: Hamiltonian})
is Hermitian \cite{key-50}. In other words, the Hamiltonian operator
(\ref{eq: Hamiltonian}) must satisfy for arbitrary wave functions
$X_{1}$ and $X_{2}$ the equality \cite{key-50}
\begin{equation}
\int dx^{3}\left[HX_{2}\right]^{*}X_{1}=\int dx^{3}X_{2}^{*}HX_{1}.\label{eq: hermiticit=0000E0}
\end{equation}
One notes that both $\vec{p}$ and $r$ are Hermitian operators. Thus,
the Hamiltonian (\ref{eq: Hamiltonian}) will automatically be a Hermitian
operator if it is a sum of a kinetic and a potential energy \cite{key-50}
\begin{equation}
H=T+V.\label{eq: energia totale}
\end{equation}
This is always the case for non-relativistic particles in Cartesian
coordinates and works also for BHs. 

Following Rosen \cite{key-40}, one can also find the BH ``Bohr radius'',
the BH wave function and the BH expected radial distance. Such quantities
and properties will be very useful in the next Section. Considering
Bohr's semi-classical model of hydrogen atom, the Bohr radius, that
is, the classical radius of the electron at the ground state, is \cite{key-40}
\begin{equation}
Bohr\:radius=\frac{1}{m_{e}e^{2}},\label{eq: Bohr radius}
\end{equation}
being $m_{e}$ the electron mass. As one wants to obtain the corresponding
``Bohr radius'' for the quantum BH, one must replace both $m_{e}$
and $e$ in Eq. (\ref{eq: Bohr radius}) with the effective mass of
the BH ground state, which is $\frac{M_{1}}{2}=1.$ Thus, the BH ``Bohr
radius'' reads
\begin{equation}
b_{1}=1,\label{eq: Bohr radius-1}
\end{equation}
which in standard units becomes $b_{1}=l_{P},$ being $l_{P}=1,61625\text{\texttimes}10^{-35}m$
the Planck length. Hence, the ``Bohr radius'' associated to the
smallest possible BH is equal to the Planck length. One finds the
wave-function associated to the BH ground state as 
\begin{equation}
\Psi_{1}=2b_{1}^{-\frac{3}{2}}r\exp-\left(\frac{r}{b_{1}}\right)=2r\exp-\left(r\right),\label{eq: wave-function 1 BH}
\end{equation}
being $\Psi_{1}$ is normalized as
\begin{equation}
\int_{0}^{\infty}\Psi_{1}^{2}dr=1.\label{eq: normalizzazione BH}
\end{equation}
The expected radial distance of this BH ground state is of the order
of 
\begin{equation}
\bar{r}_{1}=\int_{0}^{\infty}\Psi_{1}^{2}rdr=\frac{3}{2}b_{1}=\frac{3}{2}.\label{eq: size BH 1}
\end{equation}
 In fact, one has to recall that in quantum mechanics the Bohr radius
(\ref{eq: Bohr radius-1}) represents the radius having the maximum
radial probability density instead of its expected radial distance
\cite{key-51}. The latter is indeed $1.5$ times the Bohr radius
\cite{key-51}. This is due to the long tail of the radial wave function
\cite{key-51} and it is, in turn, given by Eq. (\ref{eq: size BH 1}).
One finds the expected radial distance for the BH excited at the level
$n$ as 
\begin{equation}
\bar{r}_{n}=\frac{3}{2}\left(M_{E}\right)_{n}=\frac{3}{2}\sqrt{n}.\label{eq: size BH n}
\end{equation}
In Bohr's semi-classical model of hydrogen atom, the Bohr radius for
the electron excited at the level $n$ is 
\begin{equation}
Bohr\:radius_{n}=\frac{n^{2}}{m_{e}e^{2}}.\label{eq: Bohr radius-2}
\end{equation}
Hence, if one wants to obtain the corresponding ``Bohr radius''
for the quantum BH excited at the level $n$, one must replace both
$m_{e}$ and $e$ in Eq. (\ref{eq: Bohr radius-2}) with the effective
mass of the BH excited at the level $n$. Then, by using Eq. (\ref{eq:spettro effettivo}),
one obtains
\begin{equation}
b_{n}=\sqrt{n},\label{eq: Bohr radius n}
\end{equation}
that is the half of the effective gravitational radius associated
to the BH excited at the level $n$. 

Thus, it has been shown that a BH is a well defined quantum system,
which obeys Schrödinger's theory. In a certain sense it is a ``gravitational
hydrogen atom''. One also observes that studying the BH in terms
of a well defined quantum mechanical system, having an ordered, discrete
quantum spectrum, looks consistent with the unitarity of the underlying
quantum gravity theory and with the idea that information should come
out in BH evaporation. 

\section{Quantum shell, breakdown of black hole complementarity and of firewall
paradox. Information recovery}

In this Section it is shown that Eq. (\ref{eq: Schrodinger equation BH effettiva})
is the Schrödinger equation of a self-interacting massive quantum
shell, generated by matter condensing on the apparent horizon, which
concretely realizes the membrane paradigm. This important issue enables
a series of nontrivial consequences, in particular: i) BHs have neither
horizons nor singularities; ii) there are neither information loss
in BH evaporation nor BH complementarity nor firewalls.

Let us start by rewriting the potential of Eq. (\ref{eq: energia potenziale BH effettiva})
as 
\begin{equation}
V=-\frac{M_{E}^{2}}{r}=-\frac{M^{2}}{4r}=-\frac{M^{2}}{2R},\label{eq: energia potenziale membrana}
\end{equation}
where one sets $R\equiv2r.$ The potential $V=-\frac{M^{2}}{2R}$
is well known as being the self-interaction gravitational potential
of a spherical massive shell, where $R$ is its radius \cite{key-52}.
One also recalls from previous Sections of this paper and from {[}37\textendash 39{]}
that the physical interpretation of the coordinate $r$ in the Schrödinger
equation (\ref{eq: Schrodinger equation BH effettiva}) is in terms
of the oscillating effective gravitational radius. Thus, from Eq.
(\ref{eq: effective quantities absorption finali}) $R$ results to
be the real oscillating gravitational radius. On the other hand, $V=-\frac{M^{2}}{2R}$
is also the potential of a two-particle system composed of two identical
masses $M$ gravitationally interacting with a relative position $2R$.
Thus, the spherical shell is physically equivalent to a two-particle
system of two identical masses, but, clearly, as the BH mass $M$
does not double, one has to consider the two identical masses $M$
as being fictitious and representing the real physical shell. Let
us recall the general problem of a two-particle system where the particles
have different masses \cite{key-47}. This is a 6-dimensional problem
which can be splitted into two 3-dimensional problems, that of a static
or free particle, and that of a particle in a static potential if
the sole interaction which is felt by the particles is their mutual
interaction depending only on their relative position \cite{key-47}.
One denotes by $m_{1}$, $m_{2}$ the masses of the particles, by
$\overrightarrow{d}_{1}$, $\overrightarrow{d}_{2}$ their positions
and by $\overrightarrow{p}_{1}$, $\overrightarrow{p}_{2}$ the respective
momenta. Being $\overrightarrow{d}_{1}-\overrightarrow{d}_{2}$ their
relative position, the Hamiltonian of the system reads \cite{key-47}
\begin{equation}
H=\frac{p_{1}^{2}}{2m_{1}}+\frac{p_{2}^{2}}{2m_{2}}+V(\overrightarrow{d}_{1}-\overrightarrow{d}_{2}).\label{eq: Hamiltonian 2 particles}
\end{equation}
One sets \cite{key-47}: 
\begin{equation}
\begin{array}{ccccc}
m_{T}=m_{1}+m_{2}, &  & \overrightarrow{D}=\frac{m_{1}\overrightarrow{d}_{1}+m_{2}\overrightarrow{d}_{2}}{m_{1}+m_{2}}, &  & \overrightarrow{p}_{T}=\overrightarrow{p}_{1}+\overrightarrow{p}_{2},\\
\\
m=\frac{m_{1}m_{2}}{m_{1}+m_{2}} &  & \overrightarrow{d}=\overrightarrow{d}_{1}-\overrightarrow{d}_{2} &  & \overrightarrow{p}=\frac{m_{1}\overrightarrow{p}_{1}+m_{2}\overrightarrow{p}_{2}}{m_{1}+m_{2}}.
\end{array}\label{eq: sets}
\end{equation}
The change of variables of Eq. (\ref{eq: sets}) is a canonical transformation
because it conserves the Poisson brackets \cite{key-47}. According
to the change of variables of Eq. (\ref{eq: sets}), the motion of
the two particles is interpreted as being the motion of two fictitious
particles: i) the \emph{center of mass}, having position $\overrightarrow{D}$,
total mass $m_{T}$ and total momentum $\overrightarrow{p}_{T}$ and,
ii) the \emph{relative particle} (which is the particle associated
with the relative motion), having position $\overrightarrow{d}$,
mass $m,$ called r\emph{educed mass}, and momentum $\overrightarrow{p}$
\cite{key-47}. The Hamiltonian of Eq. (\ref{eq: Hamiltonian 2 particles})
considered as a function of the new variables of Eq. (\ref{eq: sets})
becomes \cite{key-47}: 
\begin{equation}
H=\frac{p_{T}^{2}}{2m_{T}}+\frac{p^{2}}{2m}+V(\overrightarrow{d}).\label{eq: Hamiltonian separated}
\end{equation}
The new variables obey the same commutation relations as if they should
represent two particles of positions $\overrightarrow{D}$ and $\overrightarrow{d}$
and momenta $\overrightarrow{p}_{T}$ and $\overrightarrow{p}$ respectively
\cite{key-47}. The Hamiltonian of Eq. (\ref{eq: Hamiltonian separated})
can be considered as being the sum of two terms \cite{key-47}: 
\begin{equation}
H_{T}=\frac{p_{T}^{2}}{2m_{T}},\label{eq: Hamiltonian 1}
\end{equation}
and 
\begin{equation}
H_{m}=\frac{p^{2}}{2m}+V(\overrightarrow{d}).\label{eq: Hamiltonian 2}
\end{equation}
The term of Eq. (\ref{eq: Hamiltonian 1}) depends only on the variables
of the center of mass, while the term of Eq. (\ref{eq: Hamiltonian 2})
depends only on the variables of the relative particle. Thus, the
Schrödinger equation in the representation $\overrightarrow{D},\:\overrightarrow{d}$
is \cite{key-47}: 
\begin{equation}
\left[\left(-\frac{1}{2m_{T}}\triangle_{D}\right)+\left(-\frac{1}{2m}\triangle_{d}+V(d)\right)\right]\Psi\left(D,\:d\right)=E\Psi\left(D,\:d\right),\label{eq: Schrodinger equation two particles}
\end{equation}
being $\triangle_{\overrightarrow{D}}$ and $\triangle_{\overrightarrow{d}}$
the Laplacians relative to the coordinates $\overrightarrow{D}$ and
$\overrightarrow{d}$ respectively. Now, one observes that the BH
effective mass of Eq. (\ref{eq: effective quantities absorption finali})
is also the reduced mass of the previously introduced two-particle
system composed of two identical masses $M$: 
\begin{equation}
M_{E}=\frac{M*M}{M+M}=\frac{M}{2}\label{eq: massa ridotta}
\end{equation}
In that case, by recalling that in Schwarzschild coordinates the BH
center of mass coincides with the origin of the coordinate system,
and with the replacements 
\begin{equation}
\begin{array}{c}
m\rightarrow M_{E}\\
\\
d\rightarrow R,
\end{array}\label{eq: replacements}
\end{equation}
the Schrödinger equation (\ref{eq: Schrodinger equation two particles})
becomes 
\begin{equation}
\left(-\frac{1}{2M_{E}}\triangle_{2R}+V(2R)\right)\Psi\left(2R\right)=E\Psi\left(2R\right),\label{eq: Schrodinger equation membrana}
\end{equation}
which, by using Eq. (\ref{eq: energia potenziale membrana}) and $R=2r,$
in the representation $\overrightarrow{D}=0,\:\overrightarrow{r}$
reads 
\begin{equation}
-\frac{1}{2M_{E}}\left(\frac{\partial^{2}\Psi}{\partial r^{2}}+\frac{2}{r}\frac{\partial\Psi}{\partial r}\right)+V\Psi=E\Psi\label{eq: Schrodinger membrana ritrovata}
\end{equation}
for the s-states, and coincides with Eq. (\ref{eq: Schrodinger equation BH effettiva}),
with the sole difference that in the analysis in this Section the
wave function has been labelled with $\Psi$ rather than $X.$ 

Thus, one argues the intriguing physical interpretation of the analysis
in this Section. The quantum BH descripted in previous Sections in
terms of a ``gravitational hydrogen atom'' being composed by a particle,
the ``electron'', which interacts with a central field, the ``nucleus'',
is only a fictitious mathematical representation (in fact, perfect
central fields having infinite mass and pointlike ``charge'' do
not exist in nature and one has to determine the correct mass distribution
of the real quantum field \cite{key-47}) of the real, concrete physical
quantum BH, which results being a highly excited quantum shell, a
massive membrane generated by matter condensing on the apparent horizon,
which results to be an ``accumulation surface'' and having as radius
the oscillating gravitational radius. This quantum massive shell self-interacts
via the potential of Eq. (\ref{eq: energia potenziale membrana}).
This result, which has been obtained in rigorous way in \cite{key-39}
by quantizing the historical Oppenheimer and Snyder gravitational
collapse through Feynman's path integral approach,  is founded on
the sole assumption that the final result of the Oppenheimer and Snyder
gravitational collapse must be the Schwarzschild BH, which is a result
older than 80 years. Only general relativity and quantum mechanics
have been used in its derivation and it has been shown that this quantum
BH obeys Schrödinger's theory. 

The implications of this result are notable. Matter condenses on the
apparent horizon during the collapse without undergoing further collapse
and, consequently, with formation of neither horizons, nor singularities.
Instead, quantum collapse generates the oscillating quantum massive
membrane on the apparent horizon which is seems to be the real physical
state of the quantum BH. These BHs quantum shells are prevented from
collapsing by quantum gravity in a similar way to how atoms are prevented
from collapsing by quantum mechanics. This is consistent with the
approaches of Hawking \cite{key-24} and Vaz \cite{key-25} concerning
the BH information paradox \cite{key-20}. In fact, in 2014 Hawking
\cite{key-24} proposed that BH event horizons could not be the final
result of the gravitational collapse. He speculated that the BH event
horizon should be replaced by an \textquotedbl apparent horizon\textquotedbl{}
where infalling matter is suspended and then released, exactly like
in the current analysis. Hawking did not give a mechanism for how
this can work, which was later given by Vaz \cite{key-25}, who supported
Hawking's conclusion. Vaz indeed discussed an interesting quantum
gravitational model of dust collapse by showing that continued collapse
to a singularity can only be achieved by combining two independent
and entire solutions of the Wheeler-DeWitt equation \cite{key-25}.
He argued that such a combination is forbidden leading in a natural
way to matter condensing on the apparent horizon during quantum collapse
\cite{key-25}, which is the same result in this Section. In fact,
in this Section the same results of Hawking and Vaz have been obtained
via Schrödinger theory of BHs. This cannot be a coincidence and it
is also a suggestive issue because a general important problem in
quantum gravity is that different approaches usually give different
results. This is not the case of the current analysis, which retrieves
the same results of Hawking and Vaz. Our results is also consistent
with Einstein's idea of the localization of the particles within a
thin spherical shell \cite{key-53}. It is also important to stress
that before the works of Hawking and Vaz, Mitra \cite{key-26}, Schild,
Leiter and Robertson \cite{key-27} and other Authors (details can
be found in the recent book of Mitra \cite{key-28}) proposed various
approaches in which the final result of the stellar gravitational
collapse has to be an object having neither horizons nor singularities. 

Let us clarify an important point \cite{key-48}. The quantum description
in this Section implies that collapsing matter accumulates near the
(apparent) horizon and forms a shell. One might rather suspect that
this is a consequence of the previously cited fact that, as the dust
star is here treated as a whole system rather than a large number
of dust particles, hence reducing the number of degrees of freedom
to just the star radius prior to quantisation. Thus, reducing the
degrees of freedom to just the radius of the star is not enough to
estimate how collapsed matter is actually distributed. In fact, a
simple argument against the shell is provided by considering that
many dust particles (hence a significant fraction of the BH mass)
could already be inside the gravitational radius of the star when
the collapse starts: will they be pushed outwards? Or will they not
contribute to the quantised Bekenstein mass? Hence the question: how
solid can one take the conclusion that there is no horizon? Is the
probability of finding dust outside the gravitational radius completely
negligible and for any value of the BH mass? This problem is solved
in Vaz's approach \cite{key-25}. By analysing a non homogeneus dust
collpase, Vaz has shown that Dirac quantization of the constraints
leads to a Wheeler- DeWitt equation. For a smooth dust distribution,
this equation can be regularized on a lattice \cite{key-25}. Each
point on the lattice represents a collapsing dust shell and the final
solution (Eq. (5) in \cite{key-25}) represents collapse with support
everywhere in spacetime and for any value of the BH mass. This solution
results in dust shells condensing to the apparent horizon on both
sides. In fact, the wave-functions leading to Eq. (5) in \cite{key-25}
are well defined everywhere except at the apparent horizon, where
there is an essential singularity. Interior and exterior solutions
can be matched by deforming the integration path in the complex Ri
-plane so as to go around the essential singularity at the apparent
horizon \cite{key-25}.

One recalls from the previous Section that the average size of the
``gravitational hydrogen atom'' excited at the level $n$ is given
by the expected radial distance of Eq. (\ref{eq: size BH n}). Thus,
one finds an average size of the BH quantum shell given by 
\begin{equation}
\bar{R}_{n}=3\left(M_{E}\right)_{n}=3\sqrt{n},\label{eq: size shell n}
\end{equation}
and, using the mass spectrum of Eq. (\ref{eq: spettro massa BH finale})
one gets 
\begin{equation}
\bar{R}_{n}=\frac{3}{2}M_{n}.\label{eq: size shell from mass}
\end{equation}
By recalling that the classical fixed gravitational radius is $R_{g}=2M,$
one gets 
\begin{equation}
\bar{R}_{n}=\frac{3}{4}\left(R_{g}\right)_{n},\label{eq: size shell n-1-1}
\end{equation}
which means that the size of the quantum BH is $\frac{3}{4}$ of the
size of its classical counterpart. In analogous way, the maximum radial
probability density for the BH quantum shell is 
\begin{equation}
2b_{n}=2\sqrt{n},\label{eq: raggio Bohr membrana}
\end{equation}
that is, exactly the effective BH gravitational radius, as one intuitively
expects.

Hence, the surface of the Schwarzschild sphere is not an event horizon.
Instead it is a self-interacting quantum shell, a massive membrane
generated by matter condensing on the apparent horizon, which oscillates
around its average value of Eq. (\ref{eq: size shell n}) and with
a maximum radial probability density given by Eq. (\ref{eq: raggio Bohr membrana}).
This shell must have a physical thickness because of the generalized
uncertainty principle and quantum fluctuations \cite{key-54}. This
is consistent with various previous proposals in BH physics. In fact,
the first, important approach which attempted to understand which
are the BH degrees of freedom to be quantized was the membrane paradigm
\cite{key-55}. In this framework, an observer external to the BH
and static with respect to it sees the BH interactions with the external
environment as being completely described in terms of a fictitious
membrane which should be located close to the horizon and should have
various physical properties like temperature, viscosity and electrical
charge. In order to attempt solving the BH information paradox, in
\cite{key-56} Mathur proposed to realize the membrane paradigm by
finding real degrees of freedom just outside the BH horizon via string
theory, higher dimensional gravity theory and extra directions. In
the picture in \cite{key-56} the real degrees of freedom of the BH
are given by the rapidly oscillating solutions of Einstein equations
corresponding to the BH microstates. The result in the current paper
is stronger, because it shows that the BH is physically a real, self-interacting
oscillating quantum shell with a physical thickness. This BH quantum
shell is generated by matter which condenses on the apparent horizon.
It is obtained not by a conjectured paradigm but directly from the
gravitational collaspe of dust and seems being the real physical nature
of the BH. This has also profound implications on the conjecture of
BH complementarity \cite{key-22}, which states that the points of
view of the free falling observer and of the observer external to
the BH and static with respect to it should be complementary. Therefore,
one the one hand the information should be reflected at the BH horizon
and on the other hand it should pass through the BH horizon without
escaping. The principle of BH complementarity implies that no ideal
``superobserver'' can confirm both stories simultaneously. For the
observer external to the BH and static with respect to it, the infinite
time dilation at the BH horizon corresponds to the infinite amount
of time to reach the horizon. In that case, the stretched horizon
results as being hot and having a physical meaning. Hence, the observer
external to the BH and static with respect to it observes that infalling
information heats up the stretched BH horizon, which, in turn, re-radiates
it in terms of Hawking radiation. Consequently, BH evaporation should
be unitary. The infalling observer sees instead nothing special happening
at the BH horizon with both the observer and the information falling
to the BH singularity. The Authors of \cite{key-22} proposed that
both stories are complementary in the quantum sense. Thus, one gets
no contradiction without violation of linearity in quantum theory.
The conclusion of the approach in this paper has been instead that
the final result of the gravitational collapse is a self-interacting,
highly excited, spherically symmetric,\emph{ }massive quantum shell,
generated by matter condensing on the apparent horizon, which concretely
realizes the membrane paradigm. Thus, on the one hand, BHs have neither
horizons nor singularities. On the other hand, one argues that the
supposed BH complementarity arises from the use of the classical theory
at the Schwarzschild scale. But in this paper it has been shown that
the classical theory breaks down at the Schwarzschild scale because
the quantum approach that has been developed in this Section and in
previous one gives dramatically different results with respect to
the classical theory. Then, the observer external to the BH and static
with respect to it the same BH will continue to see the horizon as
being a membrane because all matter that ever fell towards the horizon
seems as being frozen forever just outside the horizon. But now, the
infalling observer must use quantum theory as he approaches the apparent
horizon (at the Schwarzschild scale) and his final destiny will be
to end up absorbed by the BH oscillating massive membrane which, in
turn, will change its mass and its total energy by jumping to a higher
(negative) energy level. Therefore, there is neither contradiction
nor complementarity between what the infalling observer sees and what
the observer external to the BH and static with respect to it sees.
Likewise there is no need to postulate the existence of a firewall.

Another fundamental issue is the absence of the information paradox.
Let us start by recalling what the BH information paradox is \cite{key-20}.
The correct description of the vacuum around a BH event horizon leads
to the emission of Hawking quanta in terms of vacuum polarization.
Hawking radiation arises from just above the horizon. Thus, the unique
BH behavior in general relativity implies the universality of Hawking
radiation. In Hawking's original derivation \cite{key-2}, BH radiance
was strictly thermal, featureless and independent of BH formation.
A more rigorous formulation of it, based on the modern language of
tunnelling, has instead shown that BH back reaction and energy conservation
imply a slight deviation from the strict thermality \cite{key-57}.
In any case, a consequence of Hawking radiation will be that the BH
will evaporate in very long times. This generates the BH information
paradox. In Hawking words, verbatim \cite{key-20} ``if there were
an event horizon, the outgoing state would be mixed. If the black
hole evaporated completely without leaving a remnant, as most people
believe and would be required by CPT, one would have a transition
from an initial pure state to a mixed final state and a loss of unitarity''.
In other words, the BH interior state cannot be reconstructed (with
the macroscopic exception of the BH ``hairs'', which are mass, charge
and angular momentum \cite{key-49}) from the exterior data, and,
in particular, from the final state of Hawking radiation. Consequently,
there is no unitary transformation of states in a Hilbert space which
can describe BH evaporation, which, in turn, seems to make quantum
gravity non-unitary. But in the above analysis it has been shown that
the real quantum state of the Schwarzschild BH is in terms of an object
free of horizons and singularities, that is, a self-interacting, highly
excited, spherically symmetric,\emph{ }massive quantum membrane, generated
by matter condensing on the apparent horizon. Thus, on the one hand,
the internal states of the oscillating quantum shell must be in causal
connection with distant observers, in agreement with the strong equivalence
principle which states that special relativity must hold locally for
all of the laws of physics in all of spacetime as seen by time-like
observers (see Section 2.1 of \cite{key-58}). This guarantees preservation
of the physical information. On the other hand, the absence of horizons
implies that the BHs oscillating quantum shells must radiate from
their surface like all other normal bodies, because radiation cannot
arise from pair creation from the vacuum. Therefore, there is no information
paradox.

It is also important to stimate the maximum value of the density of
the quantum membrane. If one returns to the BH mathematical descripton
in terms of a quantum system composed by a particle, the ``electron'',
which interacts through a quantum gravitational interaction with a
central field, the ``nucleus'', the Born rule \cite{key-59} and
the Copenhagen interpretation of quantum mechanics \cite{key-60}
imply that one cannot exactly localize the position of the ``electron'',
but one can only find the probability density of finding the ``electron''
at a given point which is, in turn, proportional to the square of
the magnitude of the wavefunction of the ``electron'' at that point.
Being the system ``electron-nucleus'' only a fictious rapresentation
of the physical quantum shell, this implies that one cannot exactly
localize the position of the quantum shell via the oscillating gravitational
radius, and must, in turn, use an average radius. The average radius
of the BH quantum shell is given by Eq. (\ref{eq: size shell n}).
Thus, by evoking again the generalized uncertainty principle, which
guarantees that the BH quantum shell must have a physical thickness,
at least of the order of the Planck lenght, one can compute the minimum
volume of the BH quantum shell (in Planck units) as the difference
between the volume of the sphere having radius $3\sqrt{n}+\frac{1}{2}$
and the volume of the sphere having radius $3\sqrt{n}-\frac{1}{2}.$
Thus, one gets: 
\begin{equation}
\begin{array}{c}
V_{min}=\frac{4}{3}\pi\left[\left(3\sqrt{n}+\frac{1}{2}\right)^{3}-\left(3\sqrt{n}-\frac{1}{2}\right)^{3}\right]\\
\\
=\frac{4}{3}\pi\left(27n+\frac{1}{4}\right)=36\pi n+\frac{\pi}{3}.
\end{array}\label{eq: volume guscio}
\end{equation}
On the other hand, the mass spectrum of the BH quantum shell is given
by Eq. (\ref{eq: spettro massa BH finale}). Hence, one obtains the
maximum value of the density of the BH quantum shell as 
\begin{equation}
\rho_{max}=\frac{2\sqrt{n}}{36\pi n+\frac{\pi}{3}}.\label{eq: densit=0000E0 massima}
\end{equation}
The maximum density decreases with increasing $n,$ as one intuitively
expects. Thus, the maximum density corresponds to the ground state
of the BH quantum membrane, that, for $n=1,$ is a density of 
\begin{equation}
\rho_{max}(n=1)=\frac{2}{36\pi+\frac{\pi}{3}}\simeq0.0175,\label{eq: ground state density}
\end{equation}
in Planck units. By recalling that the Planck density is roughly $10^{93}$
grams per cubic centimetre in standard units, one gets a value of
\begin{equation}
\rho_{max}(n=1)\simeq1.752*10^{91}\:grams\:per\:cubic\:centimetre\label{eq: densit=0000E0 rozza}
\end{equation}
for the density of the ground state of the BH quantum shell in standard
units, which is very high but about two order of magnitude less than
the Planck density. For large $n$ Eq. (\ref{eq: densit=0000E0 massima})
is well approximated by 
\begin{equation}
\rho_{max}\simeq\frac{1}{18\pi\sqrt{n}}.\label{eq: densit=0000E0 massima approssimata}
\end{equation}
For a BH having mass of the order of 10 solar masses Eq. (\ref{eq: spettro massa BH finale})
gives 
\begin{equation}
\sqrt{n}=\frac{10M_{\astrosun}}{2}=5M_{\astrosun}\sim\frac{10^{34}\:grams}{M_{p}}\sim5*10^{38}.\label{eq: quantum levels 10 solar masses}
\end{equation}
being $M_{\astrosun}\sim2*10^{33}\:grams$ the solar mass and $M_{p}\sim2*10^{-5}\:grams$
the Planck mass. By inserting the result of Eq. (\ref{eq: quantum levels 10 solar masses})
in Eq. (\ref{eq: densit=0000E0 massima approssimata}) one gets 
\begin{equation}
\rho_{max}(10M_{\astrosun})\sim\frac{1}{5*18\pi*10^{38}}\sim3.5*10^{-41}\label{eq: densita 10 masse solari}
\end{equation}
in Planck units and, being the Planck density roughly $10^{93}$ grams
per cubic centimetre in standard units, one finds a value of 
\begin{equation}
\rho_{max}(10M_{\astrosun})\sim3.5*10^{52}\:grams\:per\:cubic\:centimetre.\label{eq: densita 10 masse solari SU}
\end{equation}

\section{From Schrödinger to Klein-Gordon}

The attentive reader \cite{key-48} could ask if, given the analogy
with Newtonian gravity, there is any essential difference between
the BH quantum description in general relativity given here and what
one would obtain in Newtonian physics. For example, one could ask
if the spectra are the same. In our opinion this point, despite being
interesting, is outside the goal of this paper. The cited analogy
is due to the fact that the Lagrangian of Eq. (60) in \cite{key-39}
is written in the form $T-U$ where both $T$ and $U$ are Newtonian
quantities. On the other hand, both the quantities $T$ and $U$ have
been obtained by using the Einstein field equation of general relativity,
see \cite{key-39} for details. Thus, the above question can be reformulated
as ``can the quantities $T$ and $U$ be obtained via a pure Newtonian
formulation of the gravitational collapse or will one find some deviations
from their obatining via the Einstein field equation of general relativity?''
This could be an interesting point to be developed in a future paper. 

In this Section the non-relativistic Schröedinger equation (\ref{eq: Schrodinger equation BH effettiva})
is promoted to a relativistic Klein-Gordon equation. This step needs
to be explained better also in the light of the previous discussion
regarding the fact that up to now only non-relativistic quantities
have been quantized \cite{key-48}. Eqs. (\ref{eq: energia potenziale BH effettiva})
and (\ref{eq: Schrodinger equation BH effettiva}) are indeed a consequence
of general relativity. This must be compatible with special relativity
(to which general relativity reduces locally in the freely falling
frame of the dust because of the Equivalence Principle) \cite{key-48}.
This will allow to insert the formalism of special relativity within
the quantum analysis of this paper, writing an equation in covariant
form. In this way, time and space will be treated in the same way
and the d'Alembertian operator turns out to be relativistically invariant.
This treatment seems to work also for small mass BH which could, in
principle, be viewed as sorts of fundamental particles (they are represented
via particle-like mathematical equations). Thus, in this Section the
special relativistic corrections to the BH Schrödinger theory, which
has been developed in previous Sections, will be discussed. This will
permit us to obtain the BH Klein-Gordon equation and the corresponding
eigenvalues. By considering again the particle-like approach of Sections
3 and 4, the goal of this Section is to find the Klein-Gordon generalization
of the BH Schrödinger equation (\ref{eq: Schrodinger equation BH effettiva}).
To make this, following \cite{key-61} one recall that the derivation
of a wave equation for a particle of mass $m$ can start with the
relativistic dispersion relation for the free particle \cite{key-61}:
\begin{equation}
p^{\mu}p_{\mu}=g_{\mu\nu}p^{\mu}p^{\nu}=m^{2}.\label{eq: p mu}
\end{equation}
Then, in terms of the energy $E$ and the three-momentum $\vec{p}$
one gets the well known equation on the relativistic energy \cite{key-61}
\begin{equation}
E^{2}-\left(\vec{p}\right)^{2}=m^{2}.\label{eq: energia relativistica}
\end{equation}
The Principle of Minimal Electromagnetic Coupling, that is \cite{key-61}
\begin{equation}
p_{\mu}\rightarrow\pi_{\mu}=p_{\mu}-qA_{\mu},\label{eq: MEC}
\end{equation}
describes the interaction of a particle of charge $q$ with the electromagnetic
field. In Eq. (\ref{eq: MEC}) the four-vector potential A consists
of the scalar potential $\Phi$ and the vector potential $\vec{A}$
\cite{key-61}. These obey $\vec{B}=\nabla\times\vec{A}$ and $\vec{E}=-\nabla\Phi-\frac{\partial\vec{A}}{\partial t}$
\cite{key-61}. For an electron $q=-e$, being $e$ the charge on
the proton, which is assumed as being positive by convention. The
Coulomb potential due to a proton is $\Phi=\frac{e}{r}$ and $\vec{A}=0.$
Then, $E\rightarrow E+\frac{e^{2}}{r},$ where $r$ is the proton-electron
distance \cite{key-61}. Now, one replaces $\vec{p}\rightarrow-i\nabla$and
allows the resulting Klein-Gordon equation to act on a spacial function
$\psi\left(\vec{x}\right)$ \cite{key-61}
\begin{equation}
\left[E^{2}-m^{2}+2E\left(\frac{e^{2}}{r}\right)+\left(\frac{e^{2}}{r}\right)^{2}-\left(-i\nabla\right)^{2}\right]\psi\left(\vec{x}\right)=0.\label{eq: KG generale}
\end{equation}
From the historical point of view Eq. (\ref{eq: KG generale}), which
is now known as the Klein-Gordon equation in a Coulomb potential,
was originally derived by Schrödinger in his search for an equation
describing de Broglie waves. This approach has been indeed found in
Schrödinger's notebooks from late 1925, and he appears to have prepared
a manuscript applying it to the hydrogen atom. 

On the other hand, for solving the Klein-Gordon equation momentum
space is more convenient than coordinate space \cite{key-63,key-64}.
In momentum space Eq. (\ref{eq: KG generale}) becomes \cite{key-63}

\begin{equation}
\left(E\widehat{R^{2}}+e^{2}E\widehat{R}+e^{2}\right)\psi\left(p\right)=\widehat{R^{2}}\left(m^{2}+p^{2}\right)\psi\left(p\right).\label{eq: KG spazio momenti}
\end{equation}
For the s-states the distance squared and distance operators act in
momentum space as \cite{key-63} 
\begin{equation}
\begin{array}{c}
R^{2}\psi\left(p\right)=-\left(\frac{d^{2}}{dp^{2}}+\frac{2}{p}\frac{d}{dp}\right)\psi\left(p\right)\\
\\
R\psi\left(p\right)=i\left(\frac{d}{dp}+\frac{1}{p}\right)\psi\left(p\right).
\end{array}\label{eq: distance operators}
\end{equation}
By inserting Eq. (\ref{eq: distance operators}) in Eq. (\ref{eq: KG spazio momenti})
one finds \cite{key-63} 
\begin{equation}
\begin{array}{c}
\left(\epsilon^{2}+p^{2}\right)\frac{d^{2}\left(p\right)}{dp^{2}}+\left(\frac{2\epsilon}{p}+2iEe^{2}+6p\right)\frac{d\psi\left(p\right)}{dp}+\\
\\
+\left(e^{4}+\frac{2iEe^{2}}{p}+6\right)\psi\left(p\right)=0,
\end{array}\label{eq: KG atomo idrogeno}
\end{equation}
where $\epsilon^{2}\equiv m^{2}-E^{2}.$ Now, following again the
analogy between the s-states of the hydrogen atom and the BH particle-like
representation in Sections 3 and 4, in order to obtain the Klein-Gordon
equation of the Schwarzschild BH one makes the following replacements
in Eq. (\ref{eq: KG atomo idrogeno}) 
\begin{equation}
\begin{array}{c}
m\rightarrow M_{E}\\
\\
e\rightarrow M_{E}\\
\\
\epsilon^{2}\rightarrow\epsilon_{BH}^{2}\equiv M_{E}^{2}-E^{2}.
\end{array}\label{eq: rimpiazzi}
\end{equation}
Thus, the BH Klein-Gordon equation reads 
\begin{equation}
\begin{array}{c}
\left(\epsilon_{BH}^{2}+p^{2}\right)\frac{d^{2}\left(p\right)}{dp^{2}}+\left(\frac{2\epsilon_{BH}}{p}+2iEM_{E}^{2}+6p\right)\frac{d\psi\left(p\right)}{dp}+\\
\\
+\left(M_{E}^{4}+\frac{2iEM_{E}^{2}}{p}+6\right)\psi\left(p\right)=0.
\end{array}\label{eq: KG BH}
\end{equation}
Following \cite{key-63}, the solution of Eq. (\ref{eq: KG BH}) is
obtained in terms of the hypergeometric function as 
\begin{equation}
\psi\left(p\right)=\frac{C}{p}\left(1+\frac{ip}{\epsilon_{BH}}\right)^{-\left(\frac{3}{2}+\mu\right)}F\left[\frac{3}{2}+\mu,\frac{1}{2}-w+\mu,\mu+1,2\left(1+\frac{ip}{\epsilon_{BH}}\right)^{-1}\right],\label{eq: Solution KG BH}
\end{equation}
where $F$ is the hypergeometric function $C$ is a normalization
constant and 
\begin{equation}
\begin{array}{ccc}
\mu\equiv\sqrt{\frac{1}{4}-M_{E}^{4}}, &  & w\equiv\frac{EM_{E}^{2}}{\epsilon_{BH}}.\end{array}\label{eq: mu e vu doppio}
\end{equation}
 and The quantization condition of the energy is 
\begin{equation}
n=w-\mu-\frac{1}{2},\label{eq: quantization BH KG}
\end{equation}
which is obtained in order to ensure that the wave function (\ref{eq: Solution KG BH})
is square integrable. Then, the hypergeometric series reduces to a
polynomial. By inserting the two quantities of Eq. (\ref{eq: mu e vu doppio})
in Eq. (\ref{eq: quantization BH KG}) and by using some algebra one
gets the energy spectrum 
\begin{equation}
E_{n}=\frac{M_{E}}{\sqrt{1+\frac{M_{E}^{4}}{n-\frac{1}{2}+\sqrt{\frac{1}{4}-M_{E}^{4}}}}}.\label{eq: energy spectrum  KG BH}
\end{equation}
One notes that the BH effective mass $M_{E}$ is not a constant, but
it is given by the spectrum of Eq. (\ref{eq:spettro effettivo}).
This implies that, as it is always $M_{E}\geq1,$ then the eigenvalues
of Eq. (\ref{eq: energy spectrum  KG BH}) are all imaginary. In order
to solve the problem one needs to go beyond the assumption of minimal
coupling of Eq. (\ref{eq: MEC}) which works in the electromagnetic
case. In fact, in the gravitational potential of Eq. (\ref{eq: energia potenziale BH effettiva})
one finds the square of the mass. Thus, one needs to couple such a
scalar interaction to the square of the mass in relativistic sense
in the equation of motion \cite{key-64}. In general, for a particle
of mass $m$ this is obtained via the replacement \cite{key-64} 

\begin{equation}
m^{2}\rightarrow m^{2}+U^{2}(r),\label{eq: NMEC}
\end{equation}
where $U(r)$ is the potential of an arbitrary scalar interaction
acting on the particle of mass $m$.  The corresponding Klein-Gordon
equation for the s-states with this non-minimal coupling is \cite{key-64} 

\begin{equation}
\left[\frac{d^{2}}{dr^{2}}+E^{2}-m^{2}-U^{2}(r)\right]\psi\left(r\right)=0.\label{eq: KG NMC}
\end{equation}
By dividing this last equation by $m^{2}$ with the replacement \cite{key-64}
\begin{equation}
r'=rm,\label{eq: rimpiazzo}
\end{equation}
one finds 
\begin{equation}
\left[\frac{d^{2}}{dr'^{2}}+\frac{E^{2}}{m^{2}}-1-\frac{U^{2}(r')}{m^{2}}\right]\psi\left(r'\right)=0.\label{eq: KG NMC r'}
\end{equation}
One specifies the interaction via the first of Eqs. (\ref{eq: rimpiazzi})
and setting \cite{key-64}
\begin{equation}
\begin{array}{c}
\frac{U^{2}(r')}{M_{E}^{2}}=W(r')=-\frac{M_{E}^{2}}{r'}\\
\\
b^{2}=1-\frac{E^{2}}{M_{E}^{2}}.
\end{array}\label{eq: doppio setting}
\end{equation}
Following the analysis in \cite{key-64} one gets 
\begin{equation}
\left[\frac{d^{2}}{dr'^{2}}-b^{2}+\frac{M_{E}^{2}}{r'}\right]\psi\left(r'\right)=0.\label{eq: KG NMC BH}
\end{equation}
Again, one replaces 
\begin{equation}
\begin{array}{c}
\xi=2br'\\
\\
c=\frac{M_{E}^{2}}{2b},
\end{array}\label{eq: doppio replacement}
\end{equation}
obtaining 
\begin{equation}
\left[\frac{d^{2}}{d\xi^{2}}-\frac{1}{4}+\frac{c}{\xi}\right]\psi\left(\xi\right)=0.\label{eq: KG ro}
\end{equation}
Now, one analyses the asymptotics $\xi\rightarrow\infty$ and $\xi\rightarrow0.$
For $\xi\rightarrow\infty$ Eq.(\ref{eq: KG ro}) becomes 
\begin{equation}
\left[\frac{d^{2}}{d\xi^{2}}-\frac{1}{4}\right]\psi\left(\xi\right)=0.\label{eq: KG ro  asintoto}
\end{equation}
Thus, one immediately obtains 
\begin{equation}
\psi\left(\xi\right)\propto\exp\left(-\frac{\xi}{2}\right).\label{eq: andamento asintotico}
\end{equation}
In similar way, for $\xi\rightarrow0$ Eq. (\ref{eq: KG ro}) reduces
to 
\begin{equation}
\left[\frac{d^{2}}{d\xi^{2}}+\frac{c}{\xi}\right]\psi\left(\xi\right)=0,\label{eq: KG ro infinitesimale}
\end{equation}
and one obtains
\begin{equation}
\psi\left(\xi\right)\propto\xi,\label{eq: andamento infinitesimale}
\end{equation}
which can be normalized.

Hence, for the solution of Eq. (\ref{eq: KG ro}) one chooses 
\begin{equation}
\psi\left(\xi\right)=N\xi F\left(\xi\right)\exp\left(-\frac{\xi}{2}\right),\label{eq: soluzione KG ro}
\end{equation}
where one has still to determine $F\left(\xi\right).$ Inserting Eq.
(\ref{eq: soluzione KG ro}) in Eq. (\ref{eq: KG ro}) one obtains
an equation analogous to Eq. (\ref{eq: Schroedinger con z})
\begin{equation}
\xi\frac{d^{2}F}{d\xi^{2}}+\left(2-\xi\right)\frac{dF}{d\xi}-\left(1-c\right)F=0,\label{eq: Kummer}
\end{equation}
which solution is again the confluent hypergeometric series 
\begin{equation}
F\left(\xi\right)=Z(1-c,2,\xi).\label{eq: hypergeometric series KG}
\end{equation}
The confluent hypergeometric series can be normalized only if its
first argument equals negative integer or zero \cite{key-64}. This
gives the quantization condition 

\begin{equation}
1-c=-n_{r}\label{eq: quantizzazione KG}
\end{equation}
with 
\begin{equation}
n_{r}=0,1,2,3,....+\infty.\label{eq: n di r}
\end{equation}
 By recalling the definition of principal quantum number \cite{key-64}
\begin{equation}
n=1+n_{r},\label{eq: pqr}
\end{equation}
one finds from Eq. (\ref{eq: quantizzazione KG}) 
\begin{equation}
E_{n}=\sqrt{1-\frac{M_{E}^{4}}{4n^{2}}}M_{E},\label{eq: spettro KG}
\end{equation}
which, using Eq. (\ref{eq: struttura fine gravitazionale}) for the
``gravitational fine structure constant'' gives 
\begin{equation}
E_{n}=\frac{\sqrt{3}}{2}\left(M_{E}\right)_{n}=\frac{\sqrt{3}}{2}\sqrt{n},\label{eq: spettro KG semplificato}
\end{equation}
where for the last passage Eq. (\ref{eq:spettro effettivo}) has been
used.  Eq. (\ref{eq: spettro KG semplificato}) represents the total
relativistic energy of the ``electron'' in the particle-like mathematical
representation which considers the BH as being an ``hydrogen atom''
composed by the ``electron'' interacting with a central field, the
``nucleus''. It is again a Bekenstein-like energy spectrum $\alpha\sqrt{n}$
with $\alpha=\frac{\sqrt{3}}{2}.$ By using Eqs. (\ref{eq: BH energy levels finale.})
and (\ref{eq: spettro massa BH finale}) Eq. (\ref{eq: spettro KG semplificato})
can be rewritten as 
\begin{equation}
E_{n}=\sqrt{n}-\frac{1}{2}\sqrt{n}+\frac{\sqrt{3}-1}{2}\sqrt{n},\label{eq: totalissima separata}
\end{equation}
where $\sqrt{n}$ is the BH effective rest mass, $-\frac{1}{2}\sqrt{n}$
is the eigenvalue of the Schrödinger equation (\ref{eq: Schrodinger equation BH effettiva})
and $\frac{\sqrt{3}-1}{2}\sqrt{n}$ are the relativistic corrections.

One notes that the energy of Eq. (\ref{eq: spettro KG semplificato})
is not the total BH relativistic energy, but only the relativistic
energy of the ``electron'' in the cited fictious particle-like mathematical
representation. In order to find the total BH relativistic energy
one has to add the rest mass of the center of mass in the decomposition
of Eq. (\ref{eq: sets}). In that case, by using Eq. (\ref{eq: spettro massa BH finale})
one has 
\begin{equation}
\left(m_{T}\right)_{n}=M_{n}+M_{n}=2M_{n}=4\sqrt{n}.\label{eq: massa totale}
\end{equation}
Thus, the total BH relativistic energy, that is, the total relativistic
energy of the self-interacting massive quantum shell that has been
decomposed via Eq. (\ref{eq: sets}), is 
\begin{equation}
\left(E_{tot}\right)_{n}=E_{n}+\left(m_{T}\right)_{n}=\left(4+\frac{\sqrt{3}}{2}\right)\sqrt{n}=\left(2+\frac{\sqrt{3}}{4}\right)M_{n}.\label{eq: energia totalissima}
\end{equation}
Thus, the relativistic energy of BH ground state, that is, $n=1$,
is $4+\frac{\sqrt{3}}{2}$ in Planck units. By recalling that, in
standard units the Planck energy is 
\begin{equation}
E_{P}\simeq1.96*10^{9}\:J,\label{eq: Planck energy}
\end{equation}
one gets the relativistic energy of BH ground state in standard units
as 
\begin{equation}
\left(E_{tot}\right)_{n=1}\simeq9.54*10^{9}\:J.\label{eq: ground state energy}
\end{equation}

Putting $\gamma\equiv2+\frac{\sqrt{3}}{4}$ and recalling that in
special relativity the gamma factor is 

\begin{equation}
\gamma=\left(1-v^{2}\right)^{-\frac{1}{2}},\label{eq: gamma}
\end{equation}
one can rewrite Eq. (\ref{eq: energia totalissima}) in the intriguing
Einsteinian form 
\[
\left(E_{tot}\right)_{n}=\gamma M_{n},
\]
that means that the relativistic energy of the BH self-interacting
massive quantum shell is exactly the relativistic energy of a free
particle having the BH rest mass and travelling with a velocity given
by the solution of the equation 
\begin{equation}
\left(1-v^{2}\right)^{-\frac{1}{2}}=2+\frac{\sqrt{3}}{4},\label{eq: gamma eguagliato}
\end{equation}
which is 
\begin{equation}
v\simeq0.911.\label{eq: BH velocity}
\end{equation}

\section{Conclusion remarks}

In this work it has been shown that the final result of the quantized
historical Oppenheimer and Snyder gravitational collapse is a self-interacting
massive quantum shell, generated by matter condensing on the apparent
horizon, which obeys a Schrödinger equation in the non-relativistic
case and a Klein-Gordon equation when relativistic corrections are
taken into due account. In fact, the Author and collaborators recently
found the Schwarzschild BH Schrödinger equation {[}37\textendash 39{]}.
In that approach, the traditional classical singularity in the BH
core has been replaced by a nonsingular two-particle system where
the two components, the ``nucleus'' and the ``electron'', strongly
interact with each other through a quantum gravitational interaction.
Thus, in this picture a BH is nothing else than the gravitational
analog of the hydrogen atom. In this paper, by following with caution
the analogy between this BH Schrödinger equation and the traditional
Schrödinger equation of the $s$ states ($l=0$) of the hydrogen atom,
the BH Schrödinger equation has been solved and discussed. The approach
also permitted us to find the quantum gravitational quantities which
are the gravitational analogous of the fine structure constant and
of the Rydberg constant. Remarkably, it has been shown that such quantities
are not constants. Instead, they are dynamical quantities having well
defined discrete spectra. In particular, the spectrum of the ``gravitational
fine structure constant'' is exactly the set of non-zero natural
numbers $\mathbb{N}-\left\{ 0\right\} .$ Therefore, a first,interesting
consequence of the results in this paper is that the BH results in
a well defined quantum gravitational system, which obeys Schrödinger's
theory: the ``gravitational hydrogen atom''. This should lead to
space-time quantization based on a quantum mechanical particle approach.

On the other hand, the potential energy in the BH Schrödinger equation
has been identified as being the gravitational energy of a spherically
symmetric shell. This permitted us to show that the Schwarzschild
BH results in a  self-interacting, highly excited, spherically symmetric,
massive quantum shell (a membrane), generated by matter condensing
on the apparent horizon, which concretely realizes the membrane paradigm.
As a consequence, the quantum BH descripted in the above terms of
a ``gravitational hydrogen atom'' is only a fictitious mathematical
representation of the real, concrete physical quantum BH, which results
being a self-interacting, highly excited quantum massive shell having
as radius the oscillating gravitational radius. Thus, a series of
nontrivial consequences emerge from this interesting result. In particular:
i) BHs have neither horizons nor singularities; ii) there is neither
information loss in BH evaporation, nor BH complementarity, nor firewall
paradox. These results are consistent with previous ones by Hawking
\cite{key-24}, Vaz \cite{key-25}, Mitra \cite{key-26}, Schild,
Leiter and Robertson \cite{key-27} and other Authors \cite{key-28}. 

Finally, the special relativistic corrections to the BH Schrödinger
equation and to the energy spectrum are obtained by finding the BH
Klein-Gordon equation and the corresponding eigenvalues. 

For the sake of completeness, we stress the phenomenological implications
of the approach in this paper \cite{key-48}. The mass quantisation
as conjectured by Bekenstein \cite{key-3} (and later by Mukhanov
\cite{key-5}) has been recovered. In addition, the approach discussed
here seems in agreement with the corpuscular picture of Dvali and
Gomez \cite{key-66} and similar to the novel approach to quantum
gravity (the Nexus Paradigm) of Marongwe \cite{key-67}, which models
the quantum BH in terms of another kind of hydrogen-like solutions.
An important point seems to be the fact that astronomical observations
of EHT {[}68-73{]} indicate that astrophysical BHs should have dark
surfaces. This seems indeed consistent with the results in this paper.
In particular, the result of Eq. (\ref{eq: size shell n-1-1}), that
the size of the quantum BH is $\frac{3}{4}$ of the size of its classical
counterpart, might shed some light on the physical origin of the dark
spot in the image of supermassive BH SgrA{*} \cite{key-74}. This
dark spot is indeed noticeably smaller than the classical BH shadow.
Further studies in this direction, together with increasingly precise
astronomical observations of astrophysical BHs, could confirm or rule
out this intriguing hypothesis. 

\section{Acknowledgements}

The Author thanks two unknown Referees for very useful comments and
suggestions.


\begin{thebibliography}{10}
\bibitem{key-1}J. D. Bekenstein, The quantum mass spectrum of the
Kerr black hole, \emph{Lett. Nuovo Cimento}, \textbf{11}, 467 (1974). 

\bibitem{key-2}S. W. Hawking, Particle creation by black holes, \emph{Commun.
Math. Phys.} \textbf{43}, 199 (1975).

\bibitem{key-3}J. D. Bekenstein, Quantum Black Holes as Atoms, \emph{in
Prodeedings of the Eight Marcel Grossmann Meeting, T. Piran and R.
Ruffini, eds.,} pp. 92-111 (World Scientific Singapore 1999). 

\bibitem{key-4}S. Hod, Bohr's Correspondence Principle and the Area
Spectrum of Quantum Black Holes., \emph{Phys. Rev. Lett. }\textbf{81},
4293 (1998).

\bibitem[5]{key-5}V. Mukhanov, Are black holes quantized?, \emph{JETP
Letters} \textbf{44}, 63 (1986). 

\bibitem[6]{key-6}J. D. Bekenstein and V. Mukhanov, Spectroscopy
of the quantum black hole, \emph{Phys. Lett. B} \textbf{360}, 7 (1995).

\bibitem[7]{key-7}S. Das, P. Ramadevi and U. A. Yajnik, Black hole
area quantization, \emph{Mod. Phys. Lett. A }\textbf{17}, 993 (2002).

\bibitem[8]{key-8}A. Barvinsky, S. Das, G. Kunstatter, Quantum Mechanics
of Charged Black Holes, \emph{Phys. Lett. B} \textbf{517}, 415 (2001).

\bibitem[9]{key-9}L. Modesto, J. W. Moffat, P. Nicolini, Black holes
in an ultraviolet complete quantum gravity, \emph{Phys. Lett. B} \textbf{695},
397 (2011). 

\bibitem[10]{key-10}C. Kiefer and T. Schmitz, Singularity avoidance
for collapsing quantum dust in the Lemaitre-Tolman-Bondi model, \emph{Phys.
Rev. D} \textbf{99}, 126010 (2019). 

\bibitem[11]{key-11}P. Hajicek and C. Kiefer, Embedding variables
in the canonical theory of gravitating shells, \emph{Nucl. Phys. B}
\textbf{603}, 531 (2001). 

\bibitem[12]{key-12}P. Hajicek and C. Kiefer, Singularity avoidance
by collapsing shells in quantum gravity, \emph{Int. J. Mod. Phys.
D} \textbf{10}, 775 (2001).

\bibitem[13]{key-13}A. Saini and D. Stojkovic, Nonlocal (but also
nonsingular) physics at the last stages of gravitational collapse,
\emph{Phys. Rev. D} \textbf{89}, 044003 (2014). 

\bibitem[14]{key-14}E. Greenwood and D. Stojkovic, Quantum gravitational
collapse: non-singularity and non-locality, \emph{J. High Energ. Phys.}
\textbf{0806}, 042 (2008). 

\bibitem[15]{key-15}J. E. Wang, E. Greenwood and D. Stojkovic, Schrodinger
formalism, black hole horizons and singularity behavior, \emph{Phys.
Rev. D} \textbf{80}, 124027 (2009).

\bibitem[16]{key-16}A. Davidson, From Planck area to graph theory:
Topologically distinct black hole microstates, \emph{Phys. Rev. D}
\textbf{100}, 081502 (2019). 

\bibitem[17]{key-17}A. Davidson, Hydrogen-like spectrum of spontaneously
created brane universes with de-Sitter ground state, \emph{Phys. Lett.
B }\textbf{780}, 29 (2018). 

\bibitem[18]{key-18}M. Maggiore, Black holes as quantum membranes,
\emph{Nucl. Phys. B} \textbf{429}, 205 (1994). 

\bibitem[19]{key-19}J. D. Bekenstein, Black Holes and Entropy, \emph{Phys.
Rev. D} \textbf{7}, 2333 (1973). 

\bibitem[20]{key-20}S. W. Hawking, Breakdown of predictability in
gravitational collapse, \emph{Phys. Rev. D }\textbf{14}, 2460 (1976). 

\bibitem[21]{key-21}G. \textquoteright t Hooft, On the quantum structure
of a black hole, \emph{Nucl. Phys. B} \textbf{256}, 727 (1985). 

\bibitem[22]{key-22}L. Susskind, L. Thorlacius and J. Uglum, The
Stretched Horizon and Black Hole Complementarity, \emph{Phys. Rev.
D} \textbf{48} 3743 (1993). 

\bibitem[23]{key-23}A. Almheiri, D. Marolf, J. Polchinski, and J.
Sully, Black holes: complementarity or firewalls?, \emph{J. High Energ.
Phys.} \textbf{2013}, 62 (2013).

\bibitem[24]{key-24}S. W. Hawking, Information Preservation and Weather
Forecasting for Black Holes, \emph{arXiv:}1401.5761 (2014). 

\bibitem[25]{key-25}C. Vaz, Black holes as Gravitational Atoms, \emph{Int.
J. Mod. Phys. D} \textbf{23}, 1441002 (2014). 

\bibitem[26]{key-26}A. Mitra, Non-occurrence of trapped surfaces
and Black Holes in spherical gravitational collapse: An abridged version,
\emph{Found. Phys. Lett. }\textbf{13}, 543 (2000). 

\bibitem[27]{key-27}R. E. Schild, D. J. Leiter and S. L. Robertson,
Observations Supporting the Existence of an Intrinsic Magnetic Moment
inside the Central Compact Object within the Quasar Q0957+561, \emph{AJ
}\textbf{132}, 420 (2006). 

\bibitem[28]{key-28}A. Mitra, The Rise and Fall of the Black Hole
Paradigm, \emph{Pan Macmillan Publishing India Pvt. Ltd. }(2021). 

\bibitem[29]{key-29}O. Lunin and S. D. Mathur, Statistical interpretation
of Bekenstein entropy for systems with a stretched horizon, \emph{Phys.
Rev. Lett. 88}, \textbf{211303} (2002). 

\bibitem[30]{key-30}O. Lunin and S. D. Mathur, AdS/CFT duality and
the black hole information paradox, Nucl. Phys. B \textbf{623}, 342
(2002).

\bibitem[31]{key-31}C. Corda, Time dependent Schrödinger equation
for black hole evaporation: no information loss, \emph{Ann. Phys.}
\textbf{353}, 71 (2015). 

\bibitem[32]{key-32}C. Corda, S. H. Hendi, R. Katebi, N. O. Schmidt,
Effective state, Hawking radiation and quasi-normal modes for Kerr
black holes, \emph{J. High Energ. Phys.} \textbf{2013}, 8 (2013).

\bibitem[33]{key-33}C. Corda, Precise model of Hawking radiation
from the tunnelling mechanism, \emph{Class. Quantum Grav.} \textbf{32,}
195007 (2015). 

\bibitem[34]{key-34}C. Corda, Quasi-Normal Modes: The \textquotedblleft Electrons\textquotedblright{}
of Black Holes as \textquotedblleft Gravitational Atoms\textquotedblright ?
Implications for the Black Hole Information Puzzle, \emph{Adv. High
En. Phys.} \textbf{2015}, 867601 (2015).

\bibitem[35]{key-35}N. Bohr, On the constitution of atoms and molecules,
\emph{Philos. Mag.} \textbf{26}, 1 (1913).

\bibitem[36]{key-36}N. Bohr, XXXVII. On the constitution of atoms
and molecules, \emph{Philos. Mag.} \textbf{26}, 476 (1913).

\bibitem[37]{key-37}C. Corda and F. Feleppa, The quantum black hole
as a gravitational hydrogen atom, \emph{Adv. Theor. Math. Phys.} (2023),
pre-print in \emph{arXiv}:1912.06478.

\bibitem[38]{key-38}C. Corda, F. Feleppa and F. Tamburini, On the
quantization of the extremal Reissner-Nordstrom black hole,\emph{
EPL }\textbf{132}, 30001 (2020).

\bibitem[39]{key-39}C. Corda, F. Feleppa, F. Tamburini and I. Licata,
Quantum oscillations in the black hole horizon, \emph{Theor. Math.
Phys.} \textbf{213}, 1632 (2022).

\bibitem[40]{key-40}N. Rosen, Quantum Mechanics of a Miniuniverse,
\emph{Int. Jour. Theor. Phys. }\textbf{32}, 1435, (1993).

\bibitem[41]{key-41}J. R. Oppenheimer and H. Snyder, On Continued
Gravitational Contraction, \emph{Phys. Rev.} \textbf{56}, 455 (1939).

\bibitem[42]{key-42}L. de Broglie, Recherches sur la théorie des
Quanta, \emph{Ann. de Physique} \textbf{(10)3}, 22 (1925).

\bibitem[43]{key-43}W. H. Press, Long Wave Trains of Gravitational
Waves from a Vibrating Black Hole, \emph{Astrophys. J.} \textbf{170},
L105 (1971).

\bibitem[44]{key-44}M. Maggiore, Physical Interpretation of the Spectrum
of Black Hole Quasinormal Modes. \emph{Phys. Rev. Lett. }\textbf{100},
141301 (2008).

\bibitem[45]{key-45}E. Spallucci, A. Smailagic, Horizons and the
wave function of Planckian quantum black holes, \emph{Phys. Lett B}
\textbf{816}, 136180 (2021).

\bibitem[46]{key-46}S. W. Hawking, The Path Integral Approach to
Quantum Gravity, \emph{in General Relativity: An Einstein Centenary
Survey, eds. S. W. Hawking and W. Israel,} (Cambridge University Press,
1979).

\bibitem[47]{key-47}A. Messiah, Quantum Mechanics, Vol. 1, \emph{North-Holland,
Amsterdam} (1961).

\bibitem[48]{key-48}Private communication with the Referees.

\bibitem[49]{key-49}C. W. Misner, K. S. Thorne and J. A. Wheeler,
Gravitation \emph{W. H. Feeman and Co.}, (1973).

\bibitem[50]{key-50}H. Kleinert, Path Integrals in Quantum Mechanics,
Statistics, Polymer Physics, and Financial Markets,\emph{ }5th edition,
\emph{World Scientific, Singapore} (2009).

\bibitem[51]{key-51}N. Zettili, Quantum Mechanics: Concepts and Applications
(2nd ed.), \emph{Chichester: Wiley} (2009).

\bibitem[52]{key-52}R. Arnowitt, S. Deser, and C. W. Misner, Gravitational-Electromagnetic
Coupling and the Classical Self-Energy Problem, \emph{Phys. Rev. }\textbf{120},
313 (1960). 

\bibitem[53]{key-53}A. Einstein, On a Stationary System With Spherical
Symmetry Consisting of Many Gravitating Masses, \emph{Ann. Math. (Second
Series)} \textbf{40}, 922 (1939). 

\bibitem[54]{key-54}M. Maggiore, Black hole complementarity and the
physical origin of the stretched horizon, \emph{Phys. Rev. D} \textbf{49},
2918 (1994). 

\bibitem[55]{key-55}K. S. Thorne, R. H. Price and D. A. Macdonald,
Black Holes: The Membrane Paradigm, \emph{Yale Univ. Press }(1986). 

\bibitem[56]{key-56}S. D. Mathur, Membrane paradigm realized?, \emph{Gen.
Rel. Grav.} \textbf{42}, 2331 (2010). 

\bibitem[57]{key-57}M. K. Parikh and F. Wilczek, Hawking Radiation
as Tunneling, \emph{Phys. Rev. Lett.} \textbf{85}, 5042 (2000).

\bibitem[58]{key-58}J. A. Wheeler and I. Ciufolini, Gravitation and
inertia, \emph{Princeton University Press }(1995).

\bibitem[59]{key-59}M. Born, Zur Quantenmechanik der Stoßvorgänge,\emph{
Zeit. Phys.} \textbf{37}, 863 (1926).

\bibitem[60]{key-60}N. Bohr, The Quantum Postulate and the Recent
Development of Atomic Theory, \emph{Nature} \textbf{121}, 580 (1928).

\bibitem[61]{key-61}R. Gilmore, Lie Groups, Physics, and Geometry,
\emph{Cambridge University Press }(2008). 

\bibitem[62]{key-62}A. Kempf et al, Hilbert space representation
of the minimal length uncertainty relation, \emph{Phys. Rev. D }\textbf{52},
1108 (1995).

\bibitem[63]{key-63}D. Bouaziz, Klein-Gordon Equation with Coulomb
Potential in the Presence of a Minimal Length, \emph{in The 9th international
conference in subatomic physics and applications}, Constantine, (Algeria)
30 Sep - 02 Oct 2013, \emph{arXiv}:1311.7405.

\bibitem[64]{key-64}W. Greiner, Relativistic Quantum Mechanics, 3rd
Edition, \emph{Springer-Verlag Berlin Heidelberg}, (Germany 2000).

\bibitem[65]{key-65}S. Mendoza and S.­ Silva, The matter Lagrangian
of an ideal fluid, \emph{Int. Jour. Geom. Meth. Mod. Phys.} \textbf{18},
2150059 (2021).

\bibitem[66]{key-66}G. Dvali, C. Gomez, Black Hole's Quantum N-Portrait,
\emph{arXiv}:1112.3359 (2011).

\bibitem[67]{key-67}S. Marongwe, Horizon scale tests of quantum gravity
using the event horizon telescope observations, \emph{Int. J. Mod.
Phys. D} \textbf{32, }2350047 (2023).

\bibitem[68]{key-68}Event Horizon Telescope Collaboration et al.,
First Sagittarius A{*} Event Horizon Telescope Results. I. The Shadow
of the Supermassive Black Hole in the Center of the Milky Way, \emph{Astrophys.
J.} \textbf{930}, L12 (2022).

\bibitem[69]{key-69}Event Horizon Telescope Collaboration et al.,
First Sagittarius A{*} Event Horizon Telescope Results. II. EHT and
Multiwavelength Observations, Data Processing, and Calibration, \emph{Astrophys.
J. }\textbf{930}, L13 (2022).

\bibitem[70]{key-70}Event Horizon Telescope Collaboration et al.,
First Sagittarius A{*} Event Horizon Telescope Results. III. Imaging
of the Galactic Center Supermassive Black Hole, \emph{Astrophys. J.}
\textbf{930}, L14 (2022).

\bibitem[71]{key-71}Event Horizon Telescope Collaboration et al.,
First Sagittarius A{*} Event Horizon Telescope Results. IV. Variability,
Morphology, and Black Hole Mass, \emph{Astrophys. J.} \textbf{930},
L15 (2022).

\bibitem[72]{key-72}Event Horizon Telescope Collaboration et al.,
First Sagittarius A{*} Event Horizon Telescope Results. V. Testing
Astrophysical Models of the Galactic Center Black Hole, \emph{Astrophys.
J.} \textbf{930}, L16 (2022).

\bibitem[73]{key-73}Event Horizon Telescope Collaboration et al.,
First Sagittarius A{*} Event Horizon Telescope Results. VI. Testing
the Black Hole Metric, \emph{Astrophys. J.} \textbf{930}, L17 (2022).

\bibitem[74]{key-74}V. I. Dokuchaev, Physical Origin of the Dark
Spot in the First Image of Supermassive Black Hole SgrA{*} , \emph{Astronomy}
\textbf{1}, 93 (2022).
\end{thebibliography}
\end{document}